\documentclass[figure]{sag00}
\Received{2022 November 8}
\Accepted{2022 December 19}
\Published{$\langle$publication date$\rangle$}
\SetRunningHead{Center for Astronomy}{Outburst Variation of Z Cam-type Dwarf Novae}

\usepackage{times}
\input{pasjsup.sty}

\begin{document}

\title{Secular Variation in the Interval of Outbursts in Z Cam-type Dwarf Novae}
\author{Tomohito Ohshima \altaffilmark{1}}
\altaffiltext{1}{Nishi-Harima Astronomical Observatory, Center for Astronomy, University of Hyogo,\\
                   407--2 Nishigaichi, Sayo-cho, Hyogo 679--5313, Japan}
\email{ohshima@nhao.jp}

\maketitle


\begin{abstract}
The secular variation in the interval of outbursts in the following six Z Cam-type dwarf novae 
(including the subtype IW And-type) is investigated: Z Cam, RX And, AH Her, HL CMa, SY Cnc, 
and WW Cet. An analysis using the $O-C$ diagram shows that the interval of outbursts
is not steady in one system. The outburst properties before standstill are the decrease in outburst interval, 
enhancement of the magnitude in quiescence, and disappearance of the long outburst. Meanwhile, several objects have at least two 
typical intervals of outbursts. These characteristics are difficult to be explained only by the variation in mass transfer from the secondary. 
\end{abstract}

\section{Introduction}

Cataclysmic variables (CVs) are close binary systems consisting 
of a white dwarf primary and a late-type secondary filling 
its Roche lobe. An accretion disk is formed around the white 
dwarf by the material
transferred from the secondary via the inner Lagrangian point 
(L1).
Dwarf novae (DNe) are a class of CVs characterized by the 
presence of dwarf nova outbursts.
A dwarf nova outburst is a transient event with an amplitude of several 
magnitudes generated on the accretion disk.
The dwarf nova outburst is considered to be caused by thermal 
instability on the disk 
[for reviews, see \cite{war95book, osa96review}); \cite{hel01book}].

Z Cam-type dwarf nova is a subgroup of DNe. It is characterized by an
intermediate state called ''standstill'' in the light variations (\cite{nij30VSrecommend, deroy32zcam}).

In general, the interval of outbursts of Z Cam-type DNe is relatively short (10--30 d). 
This high frequency of outbursts implies a high mass transfer rate onto the accretion
disk from the secondary in Z Cam-type systems.

The accretion disk of DNe shows cycles of a high (hot) state and a low (cool) state.
Meanwhile, disk instability does not occur in cases where the column density is higher than a
certain borderline on the disk. In such a disk, the accretion disk is permanently hot.
A dense disk is permanently hot. Therefore, such systems are called nova-like systems (NLs).

The mass transfer rate of Z Cam-type DNe is considered to lie near the borderline of the mass
transfer rate $\dot{M}_{\rm{crit}}$,
i.e., the intermediate systems between DNe and NLs. Therefore, Z Cam-type systems
have both phases showing a cycle of outbursts (outburst phase) and standstill.
However, a certain type of variation in mass transfer rate is required 
to transit two states.

\citet{mey83zcam} proposed that the irradiation effect of the secondary star
by the luminosity of the accretion disk enhances the mass transfer rate. 
Meanwhile, \citet{ros17ss} attempted to reproduce these without variation in mass
transfer rate.

There was a debate over the cause of DN outburst, disk instability
 theory (DI; \cite{osa74DImodel}), and enhanced mass transfer theory (EMT; \cite{bat73DN}).
The variation in disk radius as outbursts repeat was revealed to correspond to the prediction by the DI theory. The debate was resolved observationally as well 
\citep{osa13v1504cygKepler,osa13v344lyrv1504cyg,ohs14eruma}. Nevertheless,
the origin of the standstill phenomenon remains unclear with respect to the DI theory.

The variation in outburst frequency is a noteworthy problem consideringthe role of mass transfer rate in the generation of standstill. \citet{sha05zcamrecur} investigated
the outburst interval in the Z Cam-type DN systems and demonstrated it to be moderately consistent.
However, the long-term variation in outburst frequency has not been discussed fully.
The variations in the interval of outbursts ($t_{\rm{rec}}$) in Z Cam-type DNe are discussed in this paper.

\section{Data set}

The observation data used for this study was extracted from the AAVSO database \citep{AAVSO},
VSNET data \citep{VSNET}, ASAS-SN Sky Patrol data \citep{ASASSN}, ASAS3 System data
\citep{ASAS3},
and Zwicky Transient Facility data \citep{ZTF,ZTF2}. 
The span of the data used is 2450000--2459884 (approximately 27 years).

The target systems are selected from among the brightest group of Z Cam-type DN systems.
The number of known Z Cam-type systems is 327. These are listed in 
International Variable Star Index (VSX) \footnote{https://www.aavso.org/vsx/}.
Meanwhile, 44 are listed in the General Catalogue of Variable Stars (GCVS).
This significant difference is largely owing to the non-registration (in GCVS) of objects newly identified by deep sky-survey. These objects are significantly 
faint and thereby, are difficult to be monitored adequately.

The brightest six objects are selected in this study: Z Cam, 
RX And, WW Cet, HL CMa, AH Her, and SY Cnc. Although IX Vel 
is the brightest Z Cam-type object in VSX, it had not been considered
as a DN object until recently \citep{kat21ixvel}. 
The fundamental property of the six objects is summarized in Table \ref{target}.

\begin{table*}
\caption{Target List}\label{target}
\begin{center}
\begin{tabular}{ccccc} \hline
Object & $M_{\rm{max}}$ & $M_{\rm{min}}$ & Orbital Period (d) & Remarks\\
\hline
Z Cam & 10.0 & 14.5 & 0.289841 & \\
RX And &  10.3 & 14.8 & 0.209893 & \\
WW Cet & 10.4 & 15.8 & 0.17578 & \\
HL CMa & 10.7 & 15.0 & 0.216787 & IW And type \\
AH Her & 10.8 & 14.9 & 0.258116 & IW And type\\
SY Cnc & 11.0 & 14.0 & 0.3823753 & \\
\hline
\multicolumn{5}{l}{The data of the range of variation in magnitude (maximum $M_{\rm{max}}$, minimum $M_{\rm{min}}$) of variations is retrieved }\\
\multicolumn{5}{l}{from GCVS 5.1. The data of the orbital period is retrieved from RKCat \citet{RKCat}.}\\
\hline
\end{tabular}
\end{center}
\end{table*}

The outburst timings are determined based on the light curve. Therefore, the start
date of brightening should be used if observed. 
However, the gap between observations inhibits this clear estimation because the observations
are generally few in the long-term monitoring light curve. Thus, the date on halfway
of rising or the earliest date the main outburst is used in many cases. Thereby, errors of a few days occur in the determination of the start date of outbursts. 
Additionally, the conjunction with the sun makes seasonal gap of observations,
which usually includes several outbursts, is inevitable.

Therefore, observed-minus-calculated ($O-C$) diagrams are adopted in this research to investigate the long-term variations in $t_{\rm{rec}}$.

\section{Result and Analysis}
 
\subsection{Z Cam}
 Z Cam is a prototype object of Z Cam-type systems. \citet{opp98zcam} 
analyzed AAVSO observation data
and indicated the long-term variation in $t_{\rm{rec}}$. However, 
the interval in this study is based on the moving average with the 1500-day window.
Thus, the variations in $t_{\rm{rec}}$ in the shorter term (i.e., in the timescale of
months or a few years) cannot be detected.

\begin{figure*}
  \begin{center}
    \FigureFile(160mm,220mm){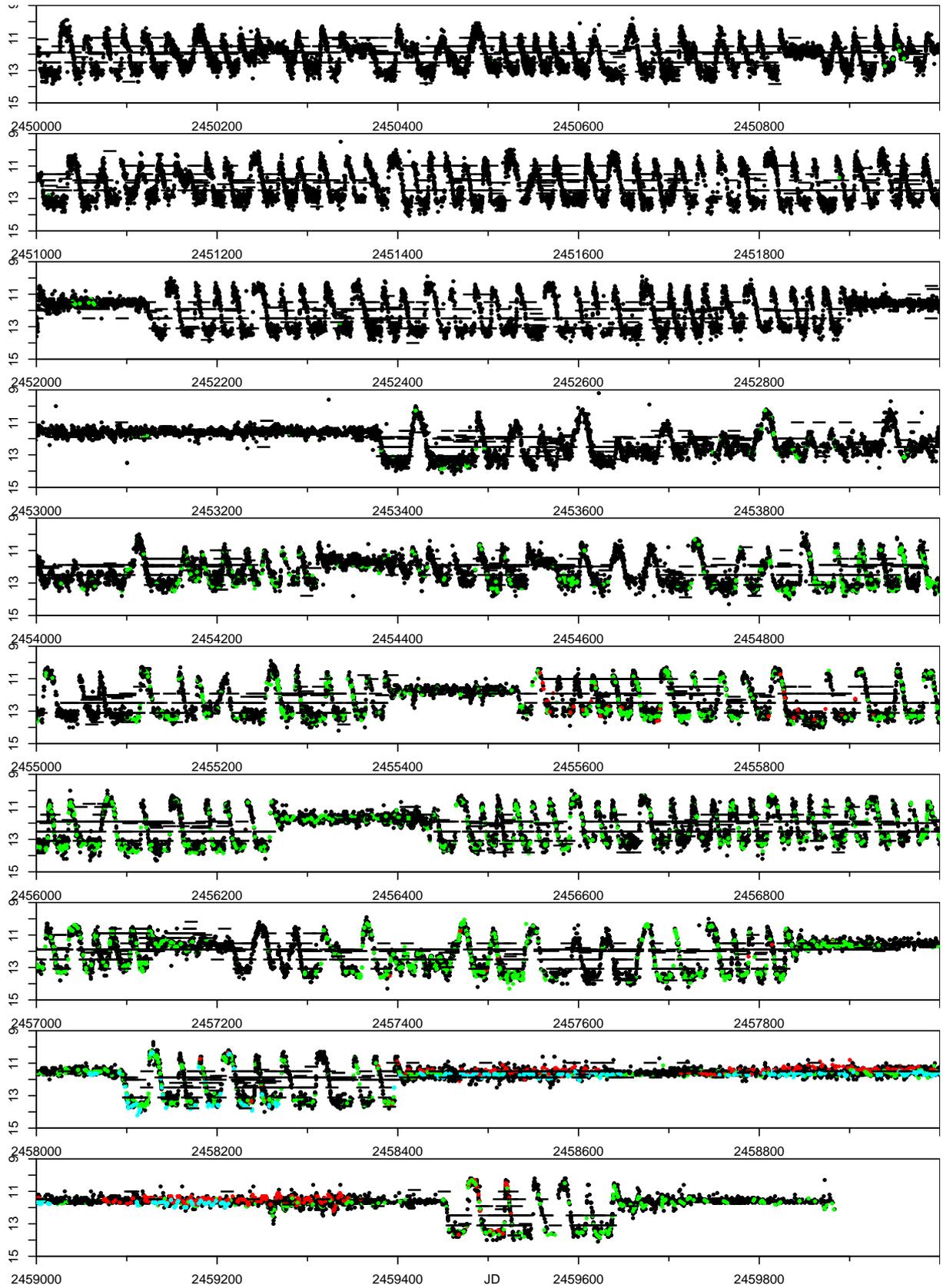}
  \end{center}
  \caption{Z Cam light curve of the entire span. The y-axis represents the magnitude. The color bands are
  as follows: the visual data (black circle), the $V$ (green circle) or CV (blue circle) 
  band on the CCD data, and the cG (red circle) band of digital camera photometry.
  The bar represents the upper limit when not observed.}\label{fig:zcamlc}
\end{figure*}

\begin{figure*}
  \begin{center}
    \FigureFile(160mm,80mm){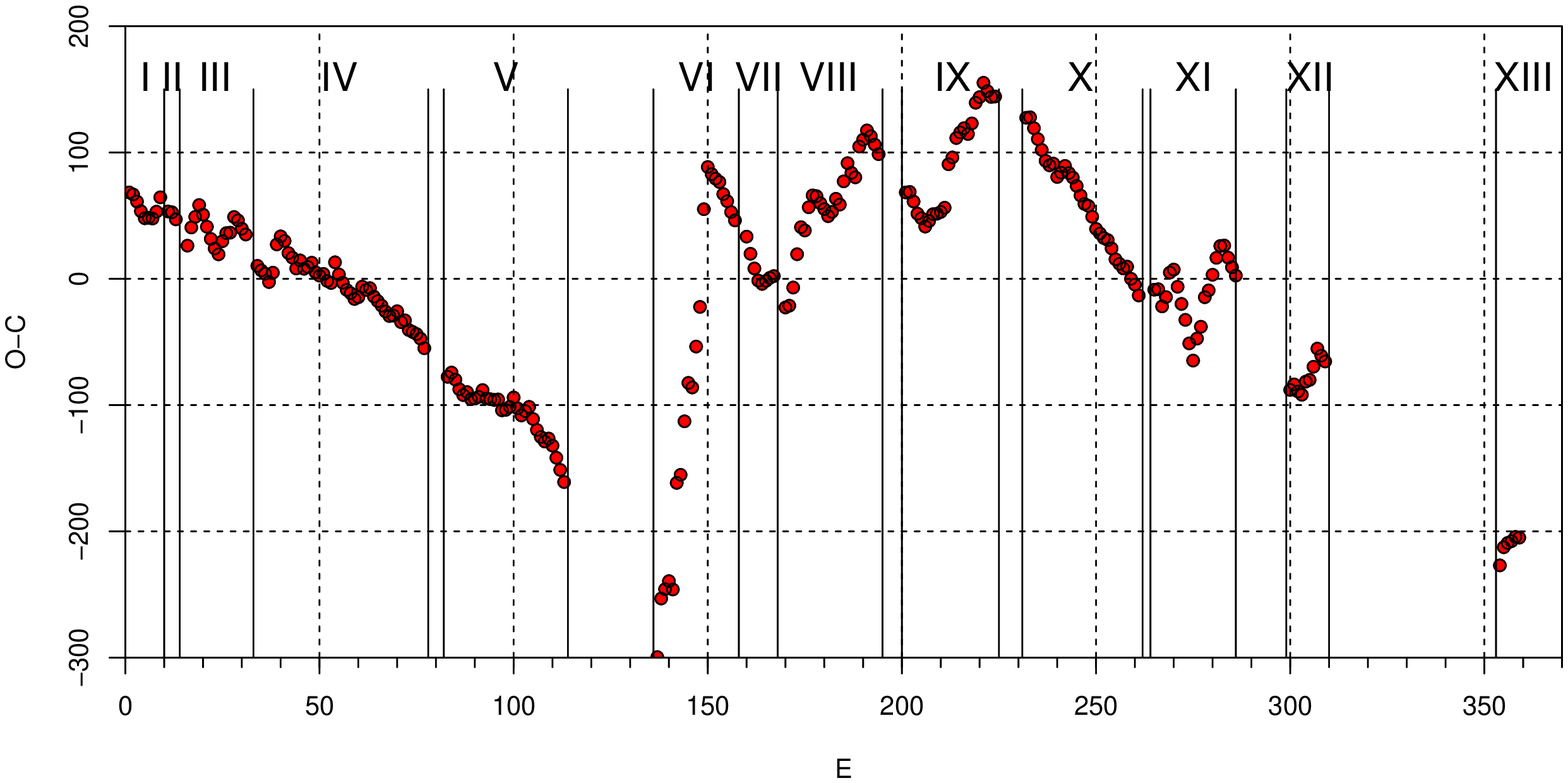}
  \end{center}
  \caption{Complete $O-C$ diagram of Z Cam. The x-axis represents the number of cycles, and the y-axis represents the
  $O-C$ value calculated with the ephemeris equation described in this paper. The Roman numbers I--XIII represent the outburst phases (divided with solid lines).}\label{fig:zcamoc}
\end{figure*}

The light curve is shown in Figure \ref{fig:zcamlc}. The Z Cam light variation broadly includes 
two phases, or standstill phases, and outbursts occur consecutively (outburst phase).
According to the light curve, the light variation is divided into 13 outburst phases and standstill phases
between these. The corresponding range for each outburst phase is plotted in Figure \ref{fig:zcamoc} (I--XIII).
Although the standstill phase is not plotted in this diagram, the journal of standstill in Z Cam is summarized
in Table \ref{zcamSS}.

Linear regression is used to estimate the mean $t_{\rm{rec}}$
 with the dates of the outbursts obtained. The period obtained 
is 27.61(6) d. 
The ephemeris JD of the outburst start date
($JD_{\rm{os}}$) is calculated based on this period by using the following ephemeris equation:

$JD_{\rm{os}} = 2449928.7 + 27.61 \times E$

The value of the $O-C$ diagram with the ephemeris calculated using this equation is presented in Figure \ref{fig:zcamoc}.

This diagram indicates the secular variation in $t_{\rm{rec}}$.
The $O-C$ diagram is divided by the standstill. The seasonal gap is insignificant because Z Cam can be observed
throughout the year in the northern hemisphere.
Rather, the $O-C$ diagram has a gap owing to a standstill. 

Because a large increase is observed in the $O-C$ diagram if 
the number of outbursts is assumed as zero, the cycle number is offset by a whole-number multiple of intervals
near the length of a standstill.


The characteristics seen outbursts before the start of standstill is widely seen;

(1) the minimum magnitude becomes brighter (typically 0.5 mag);

(2) the duration of outbursts reduces;

(3) the interval of outbursts reduces.

(1) cannot be observed in certain cases (OP VIII, XII, and XIII). This may be because of
the data scattering. (2) includes the most pronounced characteristics. 

(3) is not observed in OP VI and VII. However, these OPs are relatively peculiar. 
The outburst frequency in the earlier stage of OP VI is low
because of the circumvention of small outbursts considering the difficulty of determining their start timings.
Thus, the estimated $t_{\rm{rec}}$ is significantly long. This peculiar phase ends
in the subsequent stage of OP VI.

OP VII is also atypical. The termination timing of the standstill between OP VI and VII is ambiguous because the standstill
(when the luminosity was almost constant) smoothly transited the small oscillation. Such oscillation is generally observed
at the beginning or the final stage of the standstill \citep{szk84AAVSO}. However, in this case, the amplitude of 
oscillations increased and showed a smooth transition to the outburst of OP VII. Thus, the outbursts in the early and 
middle stages of OP VII outbursts may be interpreted as oscillations. Only the final three outbursts are 
considered to be ordinary outbursts.

The $O-C$ diagram shows that the $t_{\rm{rec}}$ of Z Cam is inconsistent. 
In addition, the interval length varies abruptly (rather than gradually) during the outburst phase.
.


Oscillations are observed at the beginning (cf: SS IV) or the final stage (cf: SS IX) of certain cases of standstill.
This is similar to the phenomenon reported in \citet{kat01zcam}.

\begin{table}
\caption{Z Cam Standstill List}\label{zcamSS}
\begin{center}
\begin{tabular}{cccc} \hline\hline
 & Period (JD--2400000) & Duration (d) \\
\hline
SS I & 50249--50275 & 26 \\
SS II & 20349--50372 & 23  \\
SS III & 50830--50868 & 38  \\
SS IV & 52012--52123 & 111   \\
SS V & 52899--53379 & 480  \\
SS VI & 54321--54380 & 59 \\
SS VII & 54552--54572 & 20   \\
SS VIII & 55393--55532 & 139   \\
SS IX & 56270--56443 & 163  \\
SS X & 57131--57217 & 86  \\
SS XI & 57840--78096 & 156   \\
SS XII & 58407--9453 & 1046 \\
SS XIII & 9650-- &  - & \\
\hline\hline
\end{tabular}
\end{center}
\end{table}

\subsection{RX And}

 RX And is one of the most popular Z Cam-type dwarf novae. It 
 is also known to have undergone an episode of a substantial decline in 1997 \citep{kat02rxand}. 
 It showed a similar (albeit significantly shorter decline) episode in 2000
\footnote{[vsnet-alert 4863], http://www.kusastro.kyoto-u.ac.jp/vsnet//Mail\\/alert4000/msg00863.html}.
 These phenomena imply that RX And has a certain mechanism 
 that causes the reduction in mass transfer rate to be nearly zero.
Hence, this system could play an important role in this study.

\begin{figure*}
  \begin{center}
    \FigureFile(160mm,220mm){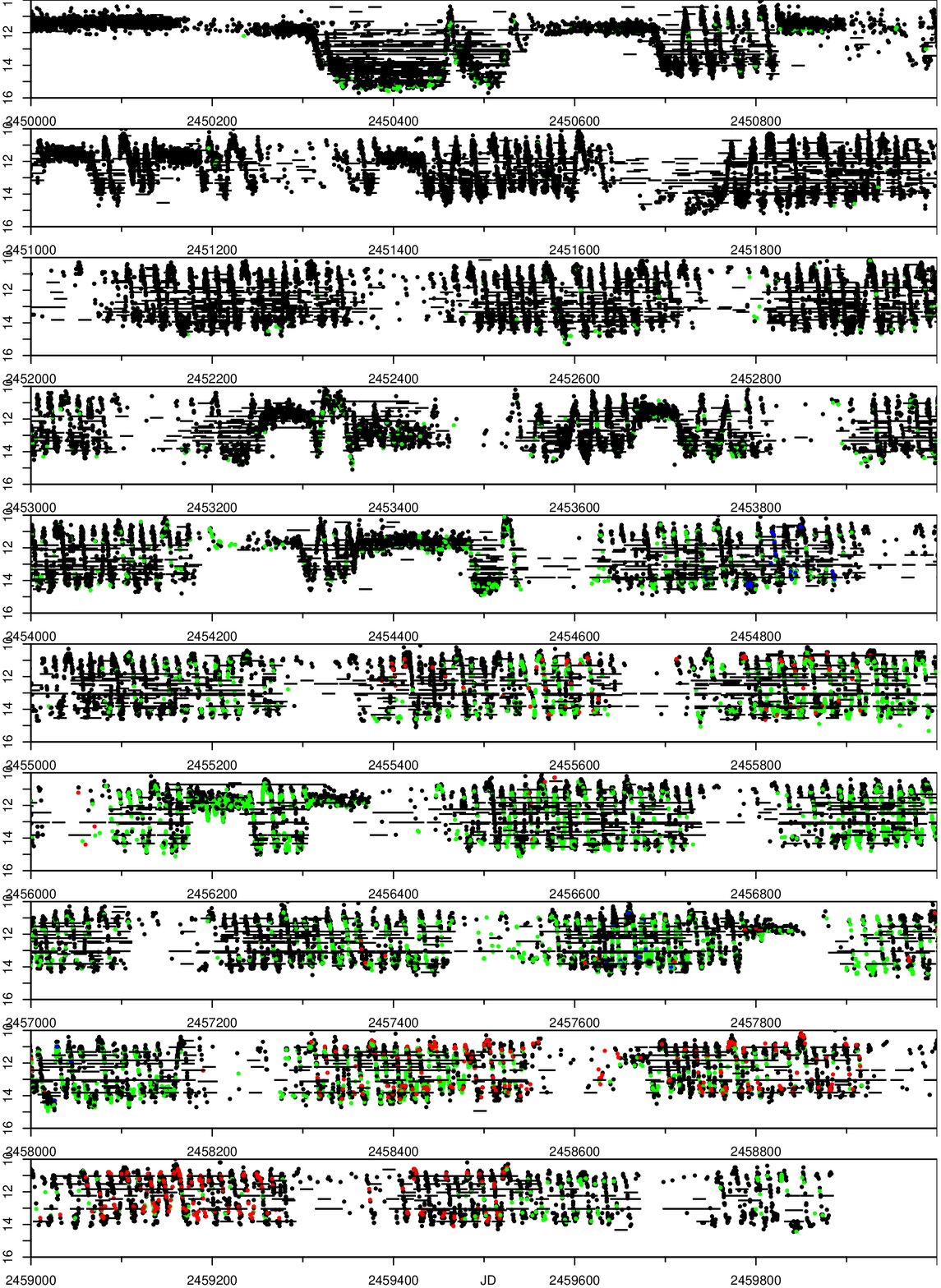}
  \end{center}
  \caption{RX And light curve of the entire span. The y-axis represents the magnitude.
  
  The color bands are
  as follows: the visual data (black circle), the $V$ (green circle) or CV (blue circle) 
  band on the CCD data, and the cG (red circle) band of digital camera photometry.
  The bar represents the upper limit when not observed.
   The bar represents the upper limit when not observed.}\label{fig:rxandlc}
\end{figure*}

The entire light curve is presented in Figure \ref{fig:rxandlc}. The data used 
starts immediately after the standstill (JD 2449900--2450300) and a 
substantial decline (JD 2450300--2450450). Ordinary variations
begin to be observed after JD 2450600. The cycle number of outbursts is
counted the standstill after that.

RX And can be observed in almost all the seasons in the northern 
hemisphere. However, the observation condition is less favorable than that for 
Z Cam because RX And is closer to the zodiac than Z Cam. 
Thus, the seasonal gap is observed more frequently than for Z Cam. 
The cycle number is estimated by the extrapolation with 
the date and the preceding interval.

The outburst phase is divided into 14 phases. Unlike One outburst
phase (OP V) is terminated by the long minimum instead a standstill.

$t_{\rm{rec}}$ is determined to be 15.239(13) d by performing linear regression on the data. \citet{szk84AAVSO} reported the mean 
period of RX And as 13 d (scattering 5--20 d).

The following is the ephemeris equation:
 
$JD_{\rm{os}} = 2450765 + 15.239 \times E$

\begin{figure*}
  \begin{center}
    \FigureFile(160mm,80mm){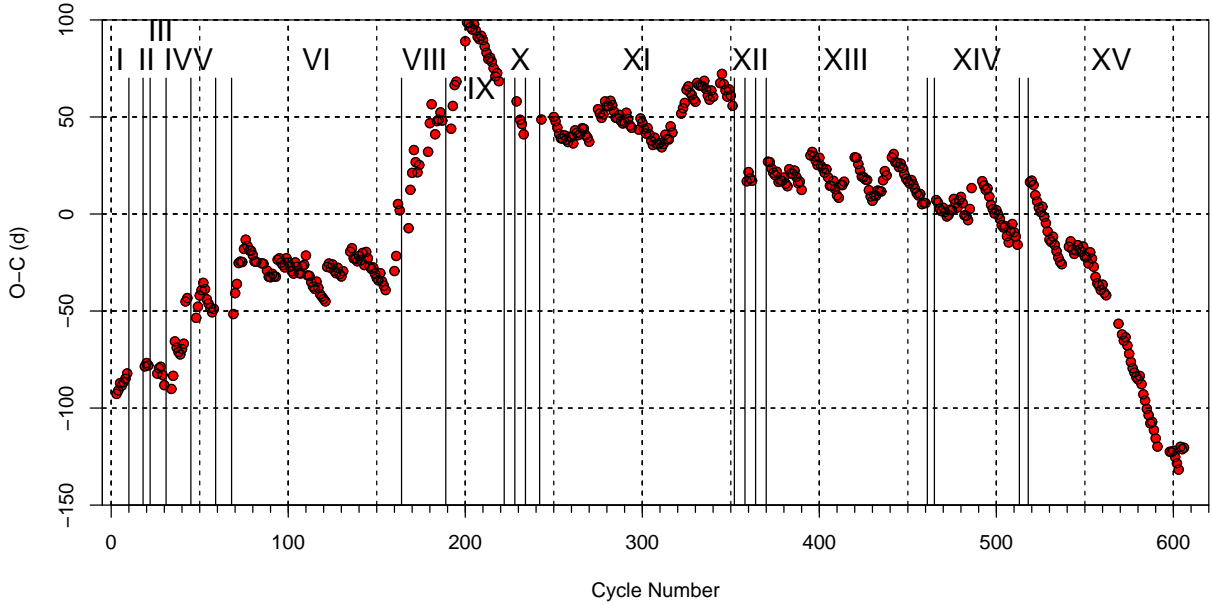}
  \end{center}
  \caption{The complete $O-C$ diagram of RX And. The x-axis represents the number of cycles, and the y-axis represents the
  $O-C$ value calculated with the ephemeris equation described in the paper. The Roman numbers I--XV represent the outburst phases (divided with solid lines).}\label{fig:rxandoc}
\end{figure*}

The $O-C$ diagram calculated with this equation is presented in figure \ref{fig:rxandoc}.
OP VIII appears to have an irregularly long interval. However, this case is
similar to OP VI of Z Cam. That is, small short outbursts are observed frequently, and
their start timings are difficult to determine.
The light curve and $O-C$ diagram reveal that the characteristics
identified in Z Cam (1)--(3) are observed in RX And as well.

Long-continued outburst phases such as OP VI, X, and XII show a cyclic variation in the $O-C$
value. This more or less corresponds to the cyclic variation in luminosity in quiescence.
However, the cyclic variation in $O-C$ occurs abruptly. Meanwhile, the variation in luminosity in quiescence occurs more gradually. 
Figure \ref{fig:rxandoc} shows two types of gradients: the downward-sloping curve and the upward-sloping curve.
This implies that the $O-C$ curve is composed of two periods: 14 d and 17 d.

\begin{table}
\caption{RX And Standstill List}\label{rxandSS}
\begin{center}
\begin{tabular}{cccc} \hline\hline
 & Period (JD - 2400000) & Duration (d) \\
\hline
SS I & 50826 -- 50965 & 139  \\
SS II & 51013 -- 51068 & 55  \\
SS III & 51135 -- 51186 & 51   \\
SS IV & 51384 -- 51431 & 47 \\
SS V & 53265 -- 53312 & 47  \\
SS VI & 53670 -- 53710 & 40  \\
SS VII & 54203 -- 54297 & 94  \\
SS VIII & 54366 -- 54485 & 119 \\
SS IX & 56179 -- 56245 & 66   \\
SS X & 56305 -- 56375* & 70  \\
SS XI & 57788 -- 57860* & 82  \\
SS XII & 58649 -- 58681 & 32  \\
\hline
\multicolumn{4}{l}{* includes errors because of observational gap.}\\
\multicolumn{4}{l}{OP V and VI are distinguished by a declining state rather than a standstill.}\\
\hline
\hline
\end{tabular}
\end{center}
\end{table}

\subsection{WW Cet}

 WW Cet is the most complex case of the six objects.
 
 This object was observed as 2.1937 Ceti by \citet{luy62wwcet}. 
 \citet{pac93wwcet} indicated that it is a member of the Z Cam 
 type. However, \citet{war87CVabsmag} and \citet{rin96wwcet} indicated 
 the unavailability of evidence to demonstrate that WW Cet shows a standstill.
However, \citet{sim11wwcet} reported the first observation of WW Cet in a standstill state. This object is a ''true'' Z Cam-type DN.

\begin{figure*}
  \begin{center}
    \FigureFile(160mm,220mm){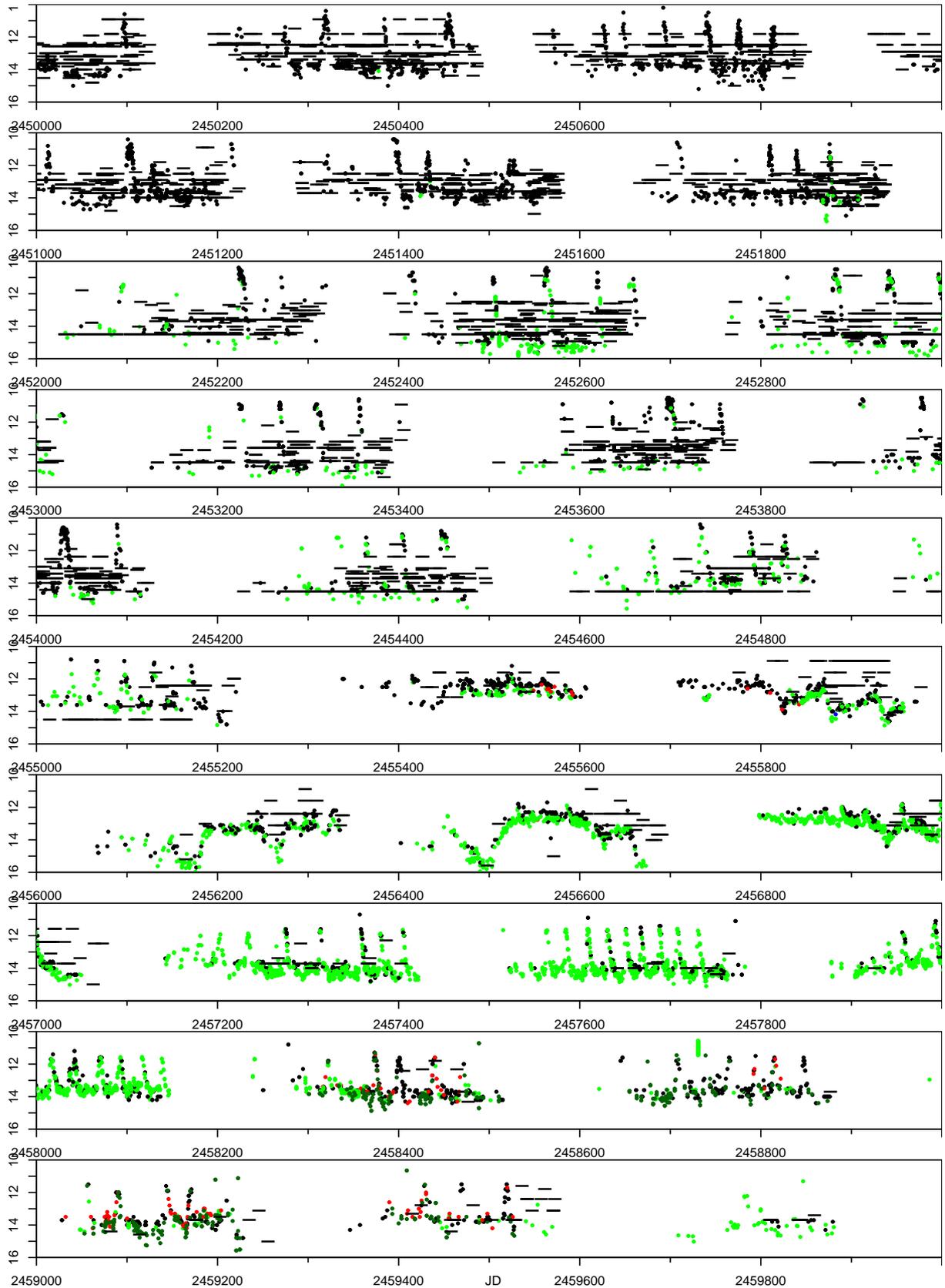}
  \end{center}
  \caption{The WW Cet light curve of the entire span. The y-axis represents the magnitude.
  The color bands are
  as follows: the visual data (black circle), the $V$ (green circle) or CV (blue circle) 
  band on the CCD data, the zg (dark green circle) band on the Zwichy data, the g (cyan circle)
  band on the ASASSN data, 
  and the cG (red circle) band of digital camera photometry.
  The bar represents the upper limit when not observed.
  The bar represents the upper limit when not observed. }\label{fig:wwcetlc}
\end{figure*}

 However, this ''standstill'' phenomenon is highly peculiar. The entire light 
 curve is shown in Figure \ref{fig:wwcetlc}, including this ''standstill''.
Although the variations of WW Cet in the 2010 
season was similar to a typical standstill, this object declined gradually
with oscillations in the 2011 season. This is unlike the behavior called standstill.
This oscillation had a period of approximately 80 ds and an amplitude of 2 magn.
This behavior is also atypical of the general DN outburst. 
Although standstills with oscillations are known, these are not accompanied by 
a gradual decline.

In addition, gradual variations during 13 mag--15 mag were observed in the 2012 and 2013 seasons.
This is also atypical of dwarf novae. 
After the typical standstill again appeared in the 2014 season, typical outbursts began to be
observed, and the general DN outburst phase started. This uncharacteristic state continued during JD 2455400--2457150 (approximately 5 years).
Therefore, the case of WW Cet cannot be regarded as the typical ''standstill''.
Apart from this uncharacteristic state, no standstill is detected in the WW Cet data used.
Therefore, the WW Cet light curve is divided into two phases: before and after the ''standstill''. 

$t_{\rm{rec}}$ is significantly long for a Z Cam-type system.
\citet{bat91wwcet} indicated that the mean $t_{\rm{rec}}$ of the WW Cet
 outburst is 30.70 d. However, $t_{\rm{rec}}$  is not identical
(ranges from 18 to 44 d).

The period is determined to be 38.4(3) d by performing linear regression on the outburst dates. With this
period, the ephemeris equation is as follows:

$JD_{\rm{os}} = 2450603 + 38.4(3) \times E$

\begin{figure*}
  \begin{center}
    \FigureFile(160mm,80mm){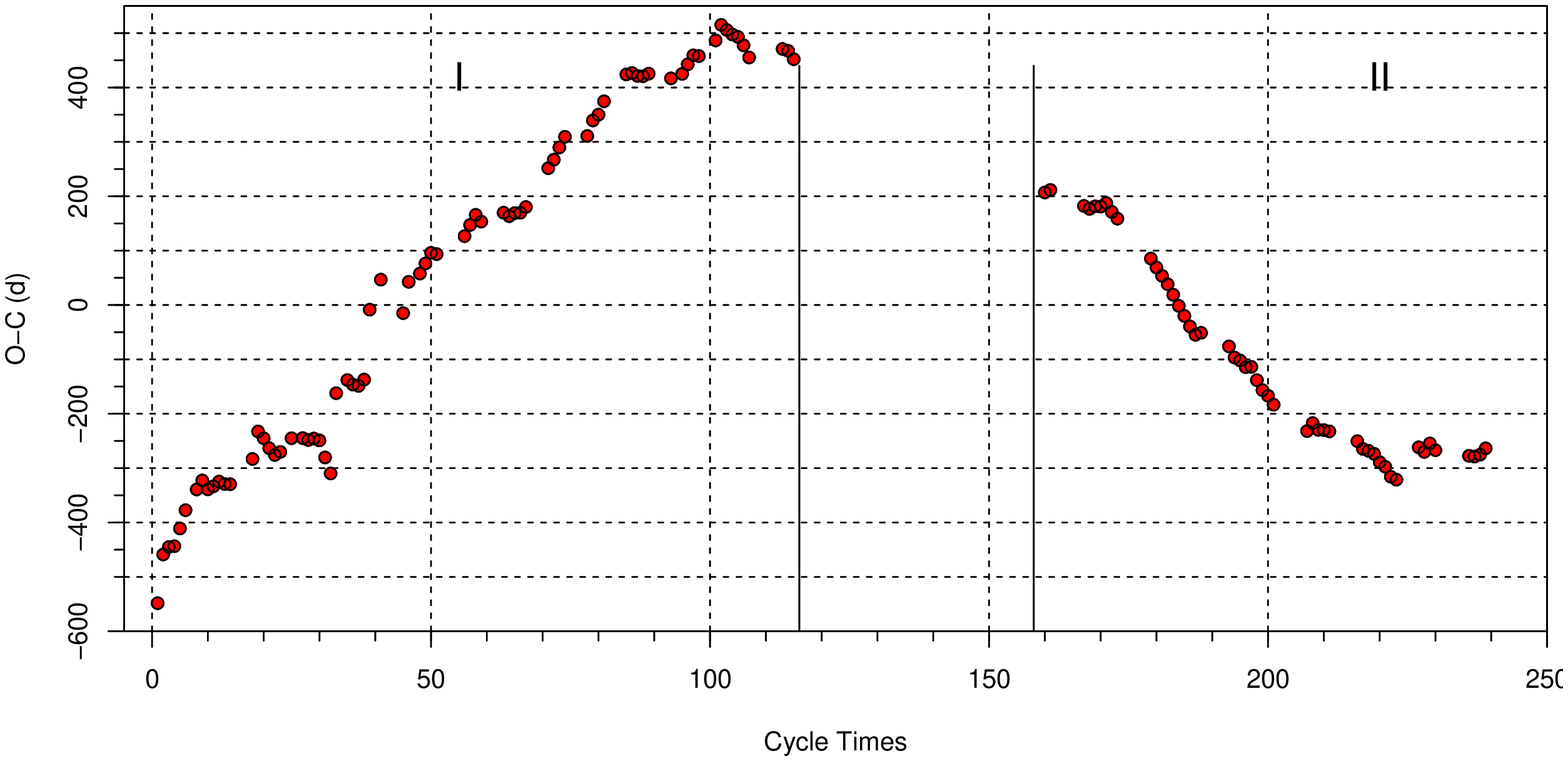}
  \end{center}
  \caption{The complete $O-C$ diagram of WW Cet. The x-axis represents the number of cycles, and the y-axis represents the
  $O-C$ value calculated with the ephemeris equation described in the paper. The Roman numbers I and II represent the outburst phases (divided with solid lines).}\label{fig:wwcetoc}
\end{figure*}

The complete $O-C$ diagram of the WW Cet outburst is shown in Figure \ref{fig:wwcetoc}. 
This diagram clearly shows the variation in $t_{\rm{rec}}$ 
because of the 
atypical ''standstill''.
The mean $t_{\rm{rec}}$ of outbursts was 48 d before the ''standstill''.
It varied to 
31 d after the ''standstill''. This is similar to the value in \citet{bat91wwcet}. 

\begin{table}
\caption{WW Cet Standstill List}
\begin{center}
\begin{tabular}{ccccc} \hline\hline
N & JD-2400000 & Duration (d) \\
\hline
SS 1 & 55300* - 56900* & 1600  \\
\hline
\multicolumn{5}{l}{* includes errors because of observational gap.}\\
\hline\hline
\end{tabular}
\end{center}
\end{table}

\subsection{HL CMa}

HL CMa was identified as an Einstein X-ray source 1E 0643.0-1648 
\citep{fuh80hlcma}. This object was identified with an optical counterpart on 
a Harvard archive plate, which showed variability with a recurrence time of 
approximately 15 d. Therefore, this object was indicated as a dwarf nova \citep{chl81hlcma}.
The classification as Z Cam-type was recommended in the context 
of the orbital period and the interval of outbursts \citep{can88outburst}. 
The existence of standstill was reported by AAVSO observations in 1982 
\citep{mau87hlcmaIUEapss}. The definite long standstill was reported in 
1999 \citep{wat00hlcma}.
\citep{kat02hlcma} reported that HL CMa shows the third state neither,
i.e. neither outburst or nor standstill (cyclic small variation).

HL CMa is recommended to be classified as a member of the new group of DNe:
IW And-type. The IW And-type object is a new subgroup of DNe similar to the Z Cam-type.
A characteristic of this type is that the standstill of this subgroup system 
terminates with the outburst \citep{kat19iwand}.
Certain members of this subgroup are classified as Z Cam-type.
\citet{sim14zcam} reported certain objects displaying this behavior in Z Cam-type
(including HL CMa).

\begin{figure*}
  \begin{center}
    \FigureFile(160mm,220mm){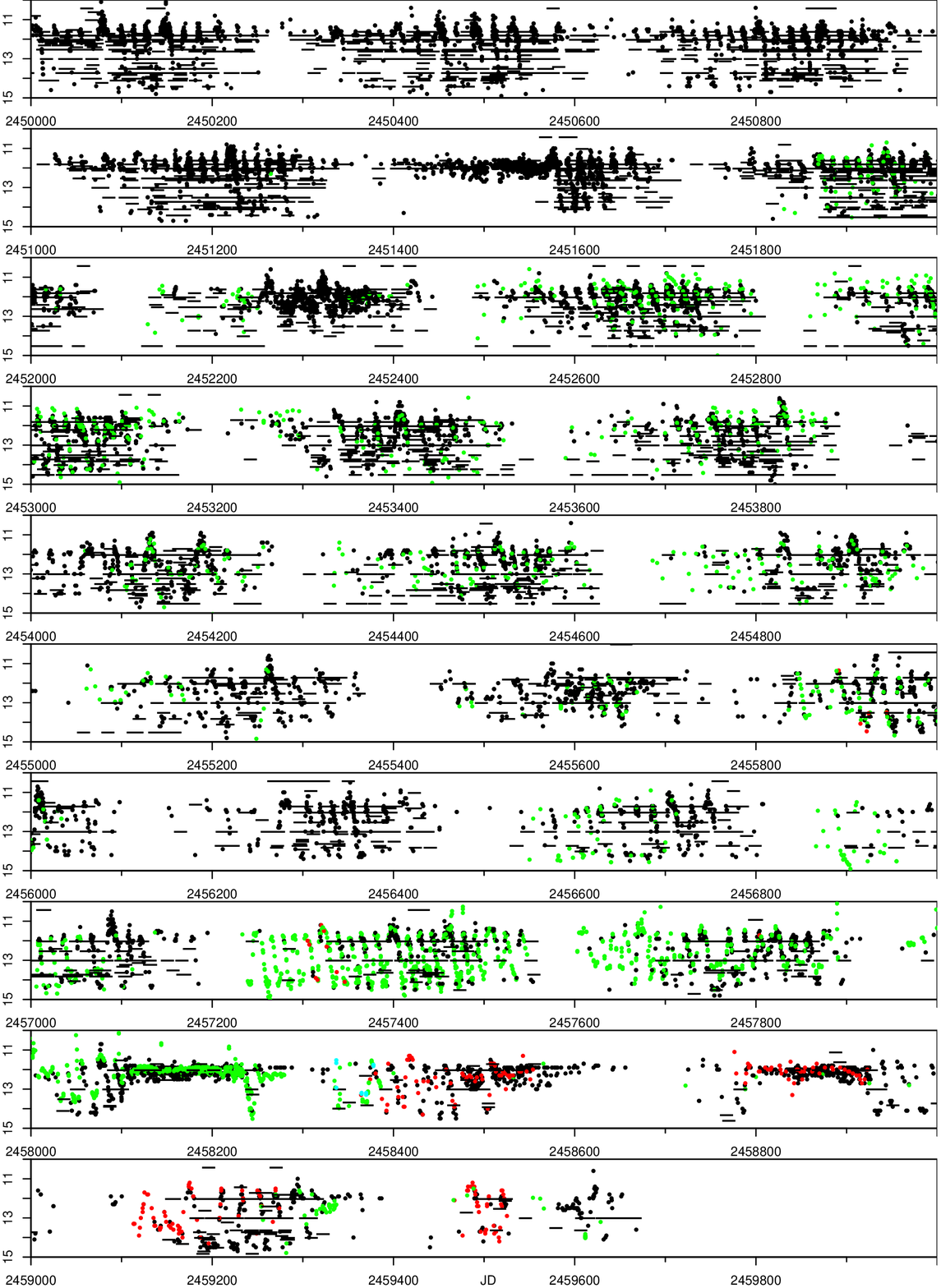}
  \end{center}
  \caption{The HL CMa light curve of the entire span. The y-axis represents the magnitude.
  The color bands are
  as follows: the visual data (black circle), the $V$ (green circle) or CV (blue circle) 
  band on the CCD data, the zg (dark green circle) band on the Zwichy data, the g (cyan circle)
  band on the ASASSN data, and the cG (red circle) band of digital camera photometry.
   The bar represents the upper limit when not observed.}\label{fig:hlcmalc}
\end{figure*}

The entire light curve is shown in Figure \ref{fig:hlcmalc}. 
According to the light curve, the light variations are divided into 11
outburst phases. These are terminated with standstill phases.
The journal of standstill in HL CMa is summarized
in Table \ref{hlcmaSS}.
Because the quiescence of HL CMa is faint (15 mag), observations 
in the minimum are highly sparse. Thus, it is challenging to discuss the variations in luminosity
in quiescence as identified in Z Cam and RX And.

The mean $t_{\rm{rec}}$ is estimated as 16.97(2) d by linear regression.
With this value, the ephemeris equation is as follows:

$JD_{\rm{os}} = 2549958 + 16.97(2)$

OP II--III are peculiar phases. These correspond to the period reported by \citet{kat02hlcma}.
Although a short ($\sim$ 60 d) standstill divides this stage into two phases, it is unclear whether this standstill (SS II) is
real or not because of the sparsity of observations of the conjunction with the sun and small oscillation.
This may be an oscillation phase with a low amplitude.

The standstill between OP VI and VII (SS VI) includes two standstills. However, only one 
outburst occurs between these two standstills. Therefore, this is not regarded as an outburst phase because an analysis
using the $O-C$ diagram cannot be performed.
This behavior wherein a few outbursts appear during standstill is generally detected in the light curve of 
IW And-type dwarf novae \citep{kat19iwand}. 
Although the light curve is unclear because of the seasonal gap, the transition of OP VII--VIII (SS VII) appears to be
a similar case.
This behavior of the light curve reveals that HL CMa shows the characteristics of IW And-type.

The $O-C$ diagram implies that HL CMa also has plural intervals of outbursts: 14--15 d and 19--20 d.
These correspond to the downward-sloping and upward-sloping parts, respectively, in the $O-C$ diagram.
\citet{sim14zcam} indicated that this object also shows IW And-type behavior.
The $O-C$ diagram implies that $t_{\rm{rec}}$ varies around $E > 350$. However, this is 
only an appearance. 


\begin{table*}
\caption{HL CMa Standstill List}\label{hlcmaSS}
\begin{center}
\begin{tabular}{cccc} \hline\hline
 & Period (JD--2400000) & Duration (d) & Remarks \\
\hline
SS I & 51412-- 51570 & 158 & TO \\
SS II & 52230*--52250* & 20 & uncertain \\
SS III & 52345--52415 & 70 & TO \\
SS IV & 53264--53310 & 46 &  \\
SS V & 55668--55690 & 22 &  \\
SS VI & 58110--58332*  & 222 & O 58246 \\
SS VII & 58475--58565*  &  90 & O 58506  \\
SS VIII & 58797--58925 & 128 &  \\
SS IX & 59310--59352 & 42 & TO \\
SS X & 59557--59603 & 46 & \\
\hline
\multicolumn{4}{l}{*:large errors because of observational gap.}\\
\multicolumn{4}{l}{TO: outburst as the termination of standstill} \\
\multicolumn{4}{l}{O: outburst during standstill and its date} \\
\hline
\hline
\end{tabular}
\end{center}
\end{table*}

\begin{figure*}
  \begin{center}
    \FigureFile(160mm,80mm){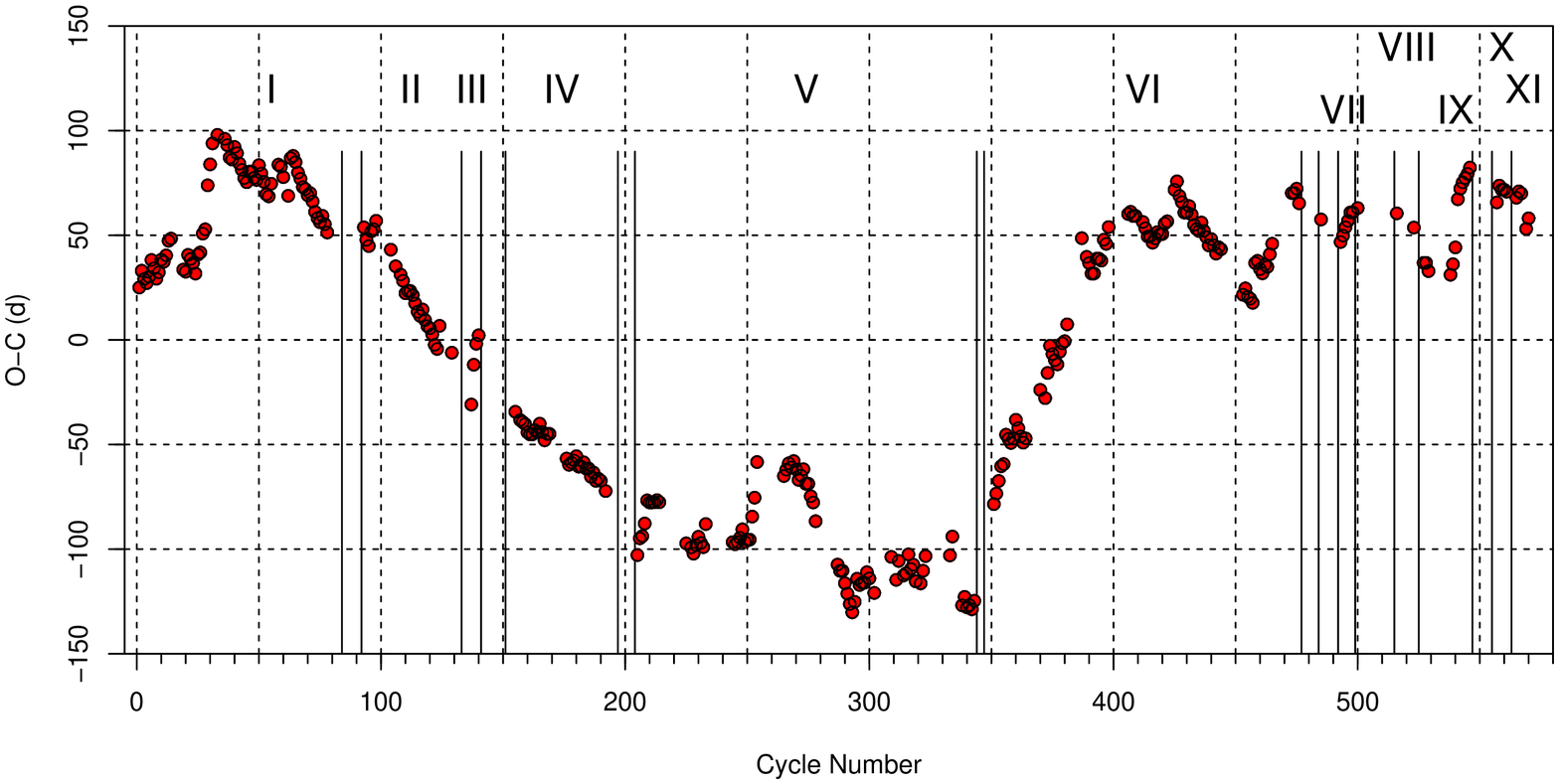}
  \end{center}
  \caption{The complete $O-C$ diagram of HL CMa. The x-axis represents the number of cycles, and the y-axis represents the
  $O-C$ value calculated with the ephemeris equation described in the paper. The Roman numbers I--XI represent the outburst phases (divided with solid lines).}\label{fig:hlcmaoc}
\end{figure*}

\subsection{AH Her}

Similar to HL CMa, AH Her is reported to display the characteristic wherein the termination of standstill
accompanies an outburst \cite{sim14zcam}.
These two systems are classified as IW And-type (UGIW) in
the VSX database.

This object was reported as a newly identified variable star in 1923 \citep{bla23ahher}. \citet{jac37ahher} indicated the relationship with Z Cam.\citet{jac37ahher} reported a period
of 19.5 d.
\citet{szk84AAVSO} estimated the interval of outbursts
as 18 d (scattering 7--27 d).

The mean $t_{\rm{rec}}$ is determined as 17.57(2) d by linear regression. With this value, the ephemeris
equation is as follows:

$JD_{\rm{os}} = 2450003 + 17.57 \times E$

\begin{figure*}
  \begin{center}
    \FigureFile(160mm,80mm){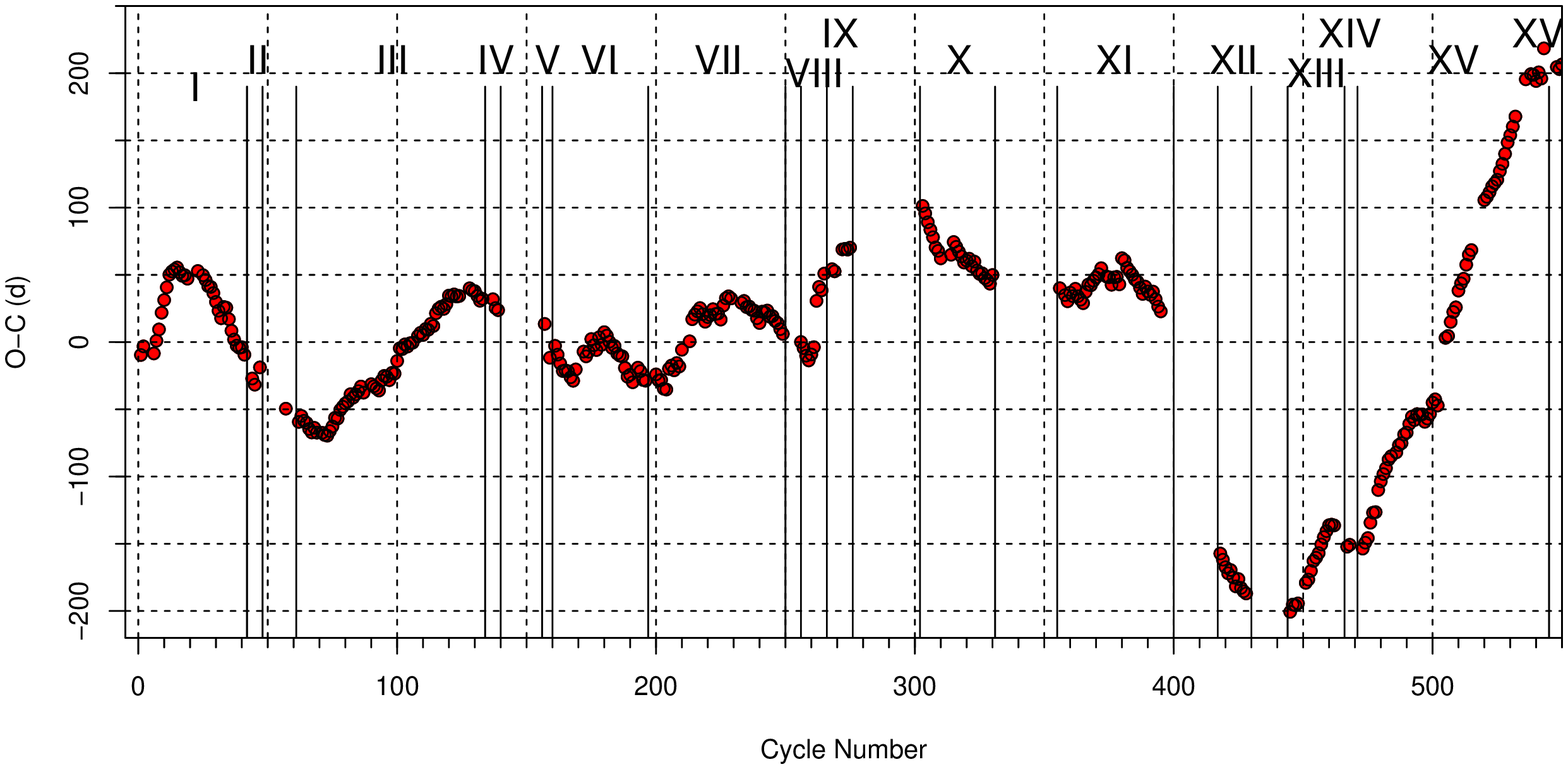}
  \end{center}
  \caption{The complete $O-C$ diagram of AH Her. The x-axis represents the number of cycles, and the y-axis represents the
  $O-C$ value calculated with the ephemeris equation described in the paper. The Roman numbers I--XV represent the outburst phases (divided with solid lines).}\label{fig:ahheroc}
\end{figure*}

The $O-C$ diagram based on this ephemeris is shown in Figure \ref{fig:ahheroc}.

The entire light curve is shown in figure \ref{fig:hlcmalc}. 
According to the light curve, the light variation is divided into 15 outburst phases (OP I--XV) and standstill phases
between these.
The journal of standstill in AH Her is summarized
in Table \ref{AHHerSS}.

Similar to HL CMa, the case wherein outbursts occur transiently during a standstill is observed.
The standstill between OP II and III includes one or two outbursts. The standstill between IV and V 
displays a similar behavior. The standstill between OP X and XI show two such transitions. 
These behaviors show that AH Her is IW And-type.

The three characteristics indicated in the Z Cam section are also identified in AH Her.
In particular, a decrease in the outburst duration is observed in many cases of OP.
The increase in minimum magnitude is observable in this object. This may be expressed as oscillations.

In AH Her, OP VII is a peculiar phase. In this phase, the minimum magnitude is significantly high, frequently attaining 12.5 mag. This is
almost as high as that in the standstill. The profile of outbursts in this OP is atypical for the Z Cam type. The increase and decrease are 
highly steep. Finally, the outburst frequency reduces, and the standstill starts. 

The early phase in OP X shows the converse behavior. The luminosity attains approximately 12.5 mag even with the maximum outburst at that time.
The luminosity in the quiescence is similar to the general OP (14 mag).

Notably, the quiescent magnitude of AH Her is generally high ($\sim$ 12.5 mag)
when the next standstill is not being approached.

\begin{figure*}
  \begin{center}
    \FigureFile(160mm,220mm){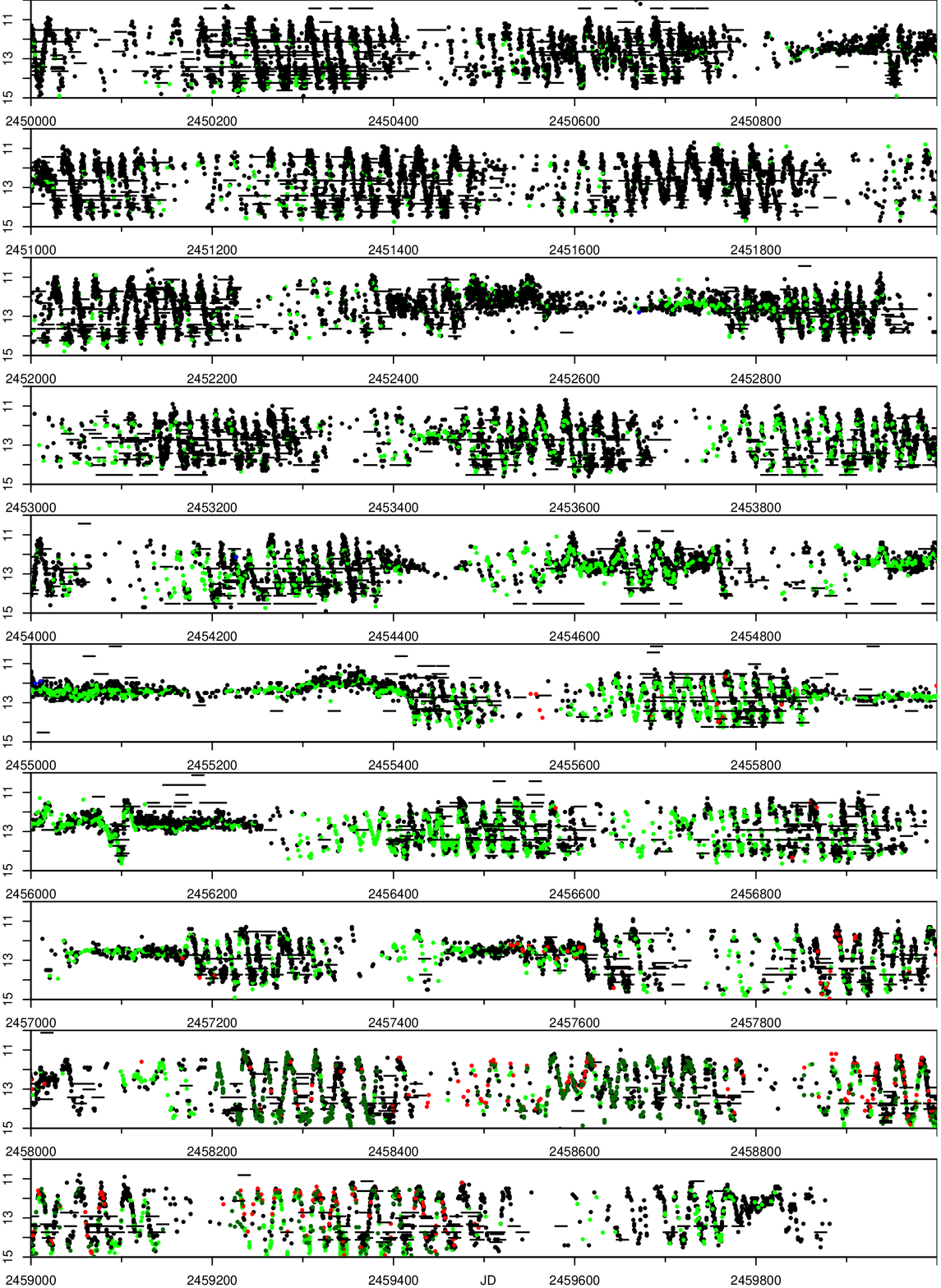}
  \end{center}
  \caption{The AH Her light curve of the entire span. The y-axis represents the magnitude. The color bands are
  as follows: the visual data (black circle), the $V$ (green circle) or CV (blue circle) 
  band on the CCD data, the zg (dark green circle) band on the Zwichy data, the g (cyan circle)
  band on the ASASSN data, and the cG (red circle) band of digital camera photometry.
   The bar represents the upper limit when not observed. }\label{fig:ahherlc}
\end{figure*}

\begin{table}
\caption{AH Her Standstill List}\label{AHHerSS}
\begin{center}
\begin{tabular}{cccc} \hline\hline
 & Period (JD--2400000) & Duration (d) & Remarks \\
\hline
SS I & 50722--50744 & 22 & \\
SS II & 50830*--51024 & 194 & O 50958 \\
SS III & 52390--52433 & 43 & \\
SS IV & 52480--54822 & 342 & O 52778 \\
SS V & 53425--53472 & 27 & TO \\
SS VI & 54391--54485 & 94 & TO \\
SS VII & 54910--55417 & 507 &  \\
SS VIII & 55860--56264 & 404 & O 56018, 56107 \\
SS IX & 57042--57169 & 127 & TO suspected \\
SS X & 57377--57608 & 43 & O 57425 \\
SS XI & 58016--58029 & 13 & TO \\
SS XII & 58100--58138 & 38 & TO \\
SS XIII & 59784--59816 & 32 & TO \\
\hline
\multicolumn{4}{l}{* includes errors because of observational gap.}\\
\multicolumn{4}{l}{TO: outburst as the termination of standstill} \\
\multicolumn{4}{l}{O: outburst during standstill and its date} \\
\hline
\hline
\end{tabular}
\end{center}
\end{table}

\subsection{SY Cnc}

SY Cnc is classified as RW Aur-type (a subgroup of the pre-main sequence star)
in the first edition of GCVS.
However, \citet{her50CVspec} revealed that the spectrum of SY
Cnc is similar to those of DNe such as SS Cyg and Z Cam. 

Although SY Cnc is classified as Z Cam-type DN, the observational
report of standstill is highly sparse. Only two short 
standstills are detected in the used data of approximately 25 years.
In addition, one of the standstills is suspect because it may be a long
outburst.
\citet{szk84AAVSO} estimated the mean $t_{\rm{rec}}$ as 27 d, and $t_{\rm{rec}}$ 
scatters during 22--35 d.

\begin{figure*}
  \begin{center}
    \FigureFile(160mm,220mm){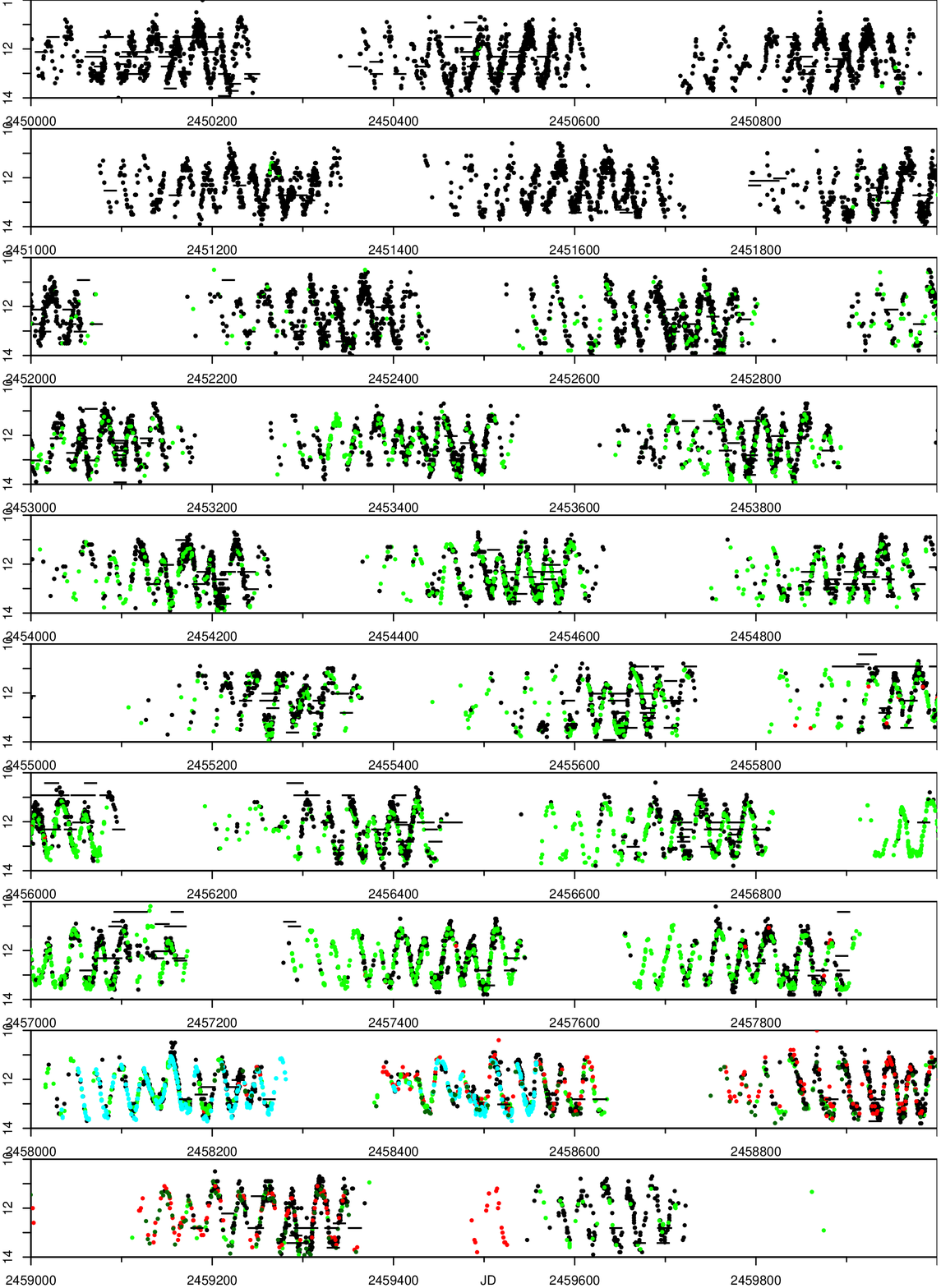}
  \end{center}
  \caption{The SY Cnc light curve of the entire span. The y-axis represents the magnitude.
  The color bands are
  as follows: the visual data (black circle), the $V$ (green circle) or CV (blue circle) 
  band on the CCD data, the zg (dark green circle) band on the Zwichy data, the g (cyan circle)
  band on the ASASSN data, and the cG (red circle) band of digital camera photometry.
  The bar represents the upper limit when not observed.}\label{fig:sycnclc}
\end{figure*}

The entire light curve is shown in Figure \ref{fig:sycnclc}. 
As described above, this object shows a standstill infrequently. 
Meanwhile, the seasonal gap is pronounced because this object 
is close to the zodiac.
$t_{\rm{rec}}$ is determined as 26.330(17) d by linear regression. This value is 
close to that presented in \citet{szk84AAVSO}.

The ephemeris equation is obtained as follows with this $t_{\rm{rec}}$:
 
$JD_{\rm{os}} = 2449887 + 26.330 \times E$

\begin{figure*}
  \begin{center}
    \FigureFile(160mm,80mm){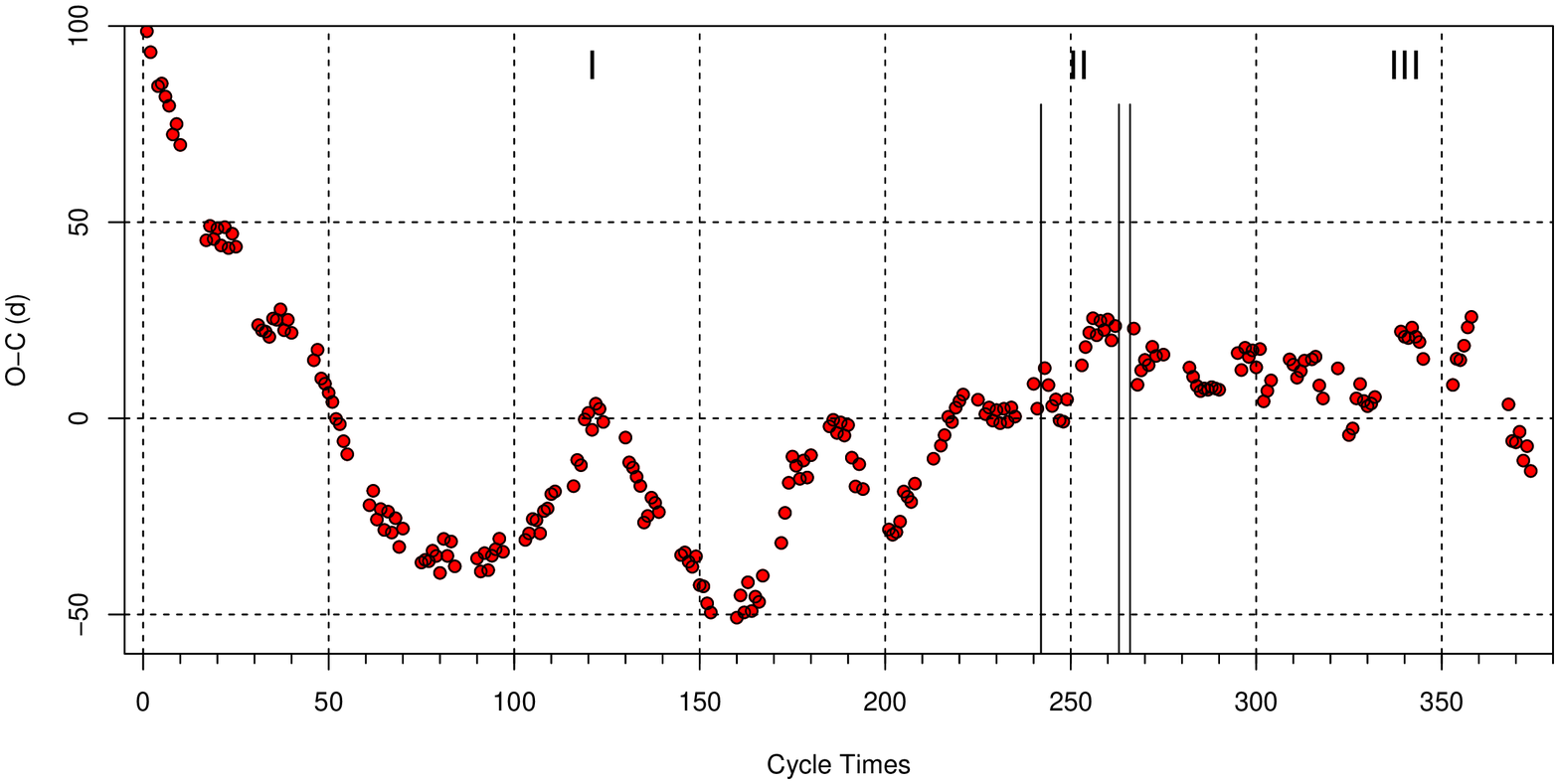}
  \end{center}
  \caption{The complete $O-C$ diagram of SY Cnc. The x-axis represents the number of cycles, and the y-axis represents the
  $O-C$ value calculated with the ephemeris equation described in the paper. The Roman numbers I--III represent the outburst phases (divided with solid lines).}\label{fig:sycncoc}
\end{figure*}

The $O-C$ diagram obtained is shown in Figure \ref{fig:sycncoc}. 
It indicates that SY Cnc displays variations in $t_{\rm{rec}}$  
regardless of the sparsity of standstill. There are two types of 
intervals. The shorter one is 23.5 d, and the longer one 
is 28 d. However, a gradual variation in $t_{\rm{rec}}$
 is observed in this object. The intermediate period is also observed.
It is nearly equal to the ''average'' length.
It is noteworthy that this variation is not related to the standstill.

\begin{table}
\caption{SY Cnc Standstill List}\label{sycncSS}
\begin{center}
\begin{tabular}{cccc} \hline\hline
 & Period (JD-2400000) &  Duration (d) & Remarks \\
\hline
SS I & 56259 - 56283 & 25 &  \\
SS II & 57110 - 57137 & 27 & uncertain \\
\hline
\hline
\end{tabular}
\end{center}
\end{table}

\section{Discussion}

The standstill frequency and durations differ among the six objects.
However, their $O-C$ diagrams indicate that all the systems show a secular variation in outburst frequency.

The trend that the 
It is notable that $t_{\rm{rec}}$ reduces as the next standstill is approached in almost all
the cases among four objects (Z Cam, RX And, AH Her, and HL CMa) except for the cases in
the highly peculiar states.
Similarly, the enhanced luminosity in the quiescence state is also interpreted
as the enhancement of the disk luminosity.

These two characteristics are explained by the variation in mass transfer rate.
Meanwhile, the long outbursts extinct before the occurrence of standstill causes
a decrease in the mass accretion from the disk. This is considered to result in
an increase in the column density of the disk and the start of the standstill.

It is noteworthy that the luminosity in quiescence is occasionally not enhanced. 
Meanwhile, the disappearance of the long outburst is observed in almost all cases.
This disappearance can be interpreted as an enhanced mass transfer
rate.

Another problem is the variation trend of $t_{\rm{rec}}$.
Certain systems (RX And and AH Her) show a cyclic variation in the quiescent luminosity during
the long-term outburst phase. Because $t_{\rm{rec}}$ shortens during bright quiescence,
this can be explained by the increased column density of the disk.
However, the $O-C$ diagrams indicate an abrupt variation in $t_{\rm{rec}}$
although the cyclic variation in quiescent luminosity is gradual.

SY Cnc shows the variation in $t_{\rm{rec}}$ during an outburst phase, particularly OP I. The abrupt variation in period 
occurs for E $\sim$ 120 (JD $sim$ 2455300) and 190 (JD $\sim$ 2454800). Both abrupt shortening of the interval.
The light curves do not show peculiarity in these timings. 

In addition, the $t_{\rm{rec}}$ in a system tends to display two values. This is difficult to explain.
SY Cnc has two typical intervals: 23.5 d and 28 d. HL CMa has two types of intervals:
14--15 d and 19--20 d. The intervals for RX And are 13 d and 17 d, and those for Z Cam are 22d and 33d.
AH Her shows more than two typical outburst intervals.

Finally, WW Cet is the most puzzling object in this research. The irregular state observed during the early 10s may have been
a true standstill. However, this may be related to the variation in mass transfer rate at any rate.
$t_{\rm{rec}}$ varied after this irregular state.

\subsection{Acknowledgement}

I acknowledge with gratitude the variable star observations obtained from the AAVSO International Database and VSNET International Database 
contributed by observers worldwide and used in this research. I also acknowledge with gratitude the data obtained by ASAS-3, ASASSN, and
The Zwicky Transient Facility. I also would like to thank Editage (www.editage.com) for English language editing.

\end{document}